\pdfoutput=1
\documentclass[11pt]{article}

\usepackage[margin=1in]{geometry}
\usepackage[T1]{fontenc}
\usepackage[utf8]{inputenc}
\usepackage{lmodern}
\usepackage{microtype}
\usepackage{amsmath,amssymb}

\usepackage{graphicx}
\usepackage{booktabs}
\usepackage{tabularx}
\usepackage{array}
\usepackage{ragged2e}
\usepackage{csquotes}
\usepackage{tikz}
\usetikzlibrary{arrows.meta,positioning,shapes.misc,shapes,fit,calc}
\usepackage{adjustbox}
\usepackage{enumitem}

\usepackage{listings}

\usepackage[numbers]{natbib}
\usepackage[hidelinks]{hyperref}
\usepackage{cleveref}

\newcommand{\Description}[1]{}

\newcommand{\keywords}[1]{%
  \par\noindent\textbf{Keywords:} #1\par
}

\title{Traceable, Enforceable, and Compensable Participation: A Participation Ledger for People-Centered AI Governance}

\author{Rashid Mushkani \\\\
  Universit\'e de Montr\'eal \\
  Mila -- Qu\'ebec AI Institute}

\date{}

\begin{document}

\maketitle

\begin{abstract}
Participatory approaches are widely invoked in AI governance, yet participation rarely translates into durable influence. In public sector and civic AI systems, community contributions such as deliberations, annotations, prompts, and incident reports are typically recorded informally, weakly linked to system updates, and disconnected from enforceable rights or sustained compensation. As a result, participation is often symbolic rather than accountable. We introduce the Participation Ledger, a machine readable and auditable framework that operationalizes participation as traceable influence, enforceable authority, and compensable labor. The ledger represents participation as an influence graph that links contributed artifacts to verified changes in AI systems, including datasets, prompts, adapters, policies, guardrails, and evaluation suites. The design integrates three core elements. First, a Participation Evidence Standard records who participated, under what consent, privacy, and compensation terms, how representational gaps were addressed, and how contributions may be reused. Second, an influence tracing mechanism ties system updates to replayable before and after tests, preserving provenance from community raised concerns to implemented changes. These tests can be re executed in future releases, forming a participatory evaluation harness that detects regressions and erosion of commitments over time. Third, the ledger encodes rights and incentives. Capability Vouchers allow authorized community stewards to request, constrain, or pause specific system capabilities within defined adoption boundaries, while Participation Credits support ongoing recognition and compensation when contributed tests continue to provide value. We ground the ledger in four deployed cases in urban AI and public space governance, spanning participatory datasets, civic planning tools, evaluation frameworks, and generative design platforms. Across cases, we identify recurring documentation and accountability gaps and show how the ledger makes participation legible, enforceable, and durable. We contribute a machine readable schema, reference templates, and an evaluation plan for assessing traceability, enforceability, and compensation in real world deployments.
\end{abstract}


\section{Introduction}

AI systems are increasingly incorporated into public decision-making and civic workflows, including tools that analyze urban environments, support consultation processes, or help stakeholders visualize future design alternatives \citep{NIST_AIRMF_2023,Canada_Directive_ADM_2019,EU_AI_Act_2024}. In urban planning, these systems shape how priorities are represented, whose values become legible, and which trade-offs are operationalized under constrained budgets, limited staff capacity, and fragmented vendor ecosystems \citep{Arnstein1969,CostanzaChock2020,Benjamin2019}. Participation is therefore a recurring governance commitment in responsible AI and public-sector frameworks, which emphasize stakeholder involvement, contestability, and documentation \citep{Reisman2018AIA,EU_AI_Act_2024,NIST_AIRMF_2023}. In practice, however, participation artifacts often remain weakly connected to technical updates and operational decisions.

This work focuses on three gaps that recur across participatory AI projects. First, participation is rarely \emph{traceable} across versions: it is difficult for an independent reviewer to connect a specific contribution to a specific system change and to the associated behavioral evidence \citep{Singh2018DecisionProvenance,Gebru2021Datasheets,Mitchell2019ModelCards}. Second, participation is rarely \emph{enforceable} within the operational boundary where a model is used: consultation outputs are commonly advisory and do not directly instantiate procedures to pause capabilities, require evaluation gates, or document remediation commitments \citep{Cobbe2021ReviewableADM,Reisman2018AIA}. Third, participation is rarely \emph{compensated} in ways that reflect ongoing maintenance value, such as repeated reuse of contributed tests or sustained harm reporting \citep{GraySuri2019GhostWork,10.1145/3706598.3713112,CostanzaChock2020}.

We address these gaps by specifying a Participation Ledger. The ledger is a machine-readable influence graph that records participation artifacts and links them to changes in AI system components and to replayable tests of change. The ledger also represents boundary-scoped rights and incentives that can be adopted in organizational policy or procurement requirements.

\paragraph{Contributions.}
\begin{itemize}
  \item We specify a Participation Ledger as an influence graph with stable identifiers and links from contributions to changes, tests of change, evaluation runs, and deployments.
  \item We define a Participation Evidence Standard that operationalizes participation reporting fields, including roles and intermediaries, recruitment pathways, consent and privacy scope, compensation terms, and reuse and retention boundaries.
  \item We introduce two governance primitives that the ledger can represent within an adoption boundary: Capability Vouchers for conditioning capability activation and Participation Credits for auditable attribution tied to replayable tests.
  \item We ground the specification in a document-based cross-case analysis of four participatory urban AI projects and report recurring evidence gaps relative to the standard, alongside a minimal feasibility validation via schema validation and an evaluation design for future deployments.
\end{itemize}

\begin{figure}[t]
  \centering
  \includegraphics[width=0.5\linewidth]{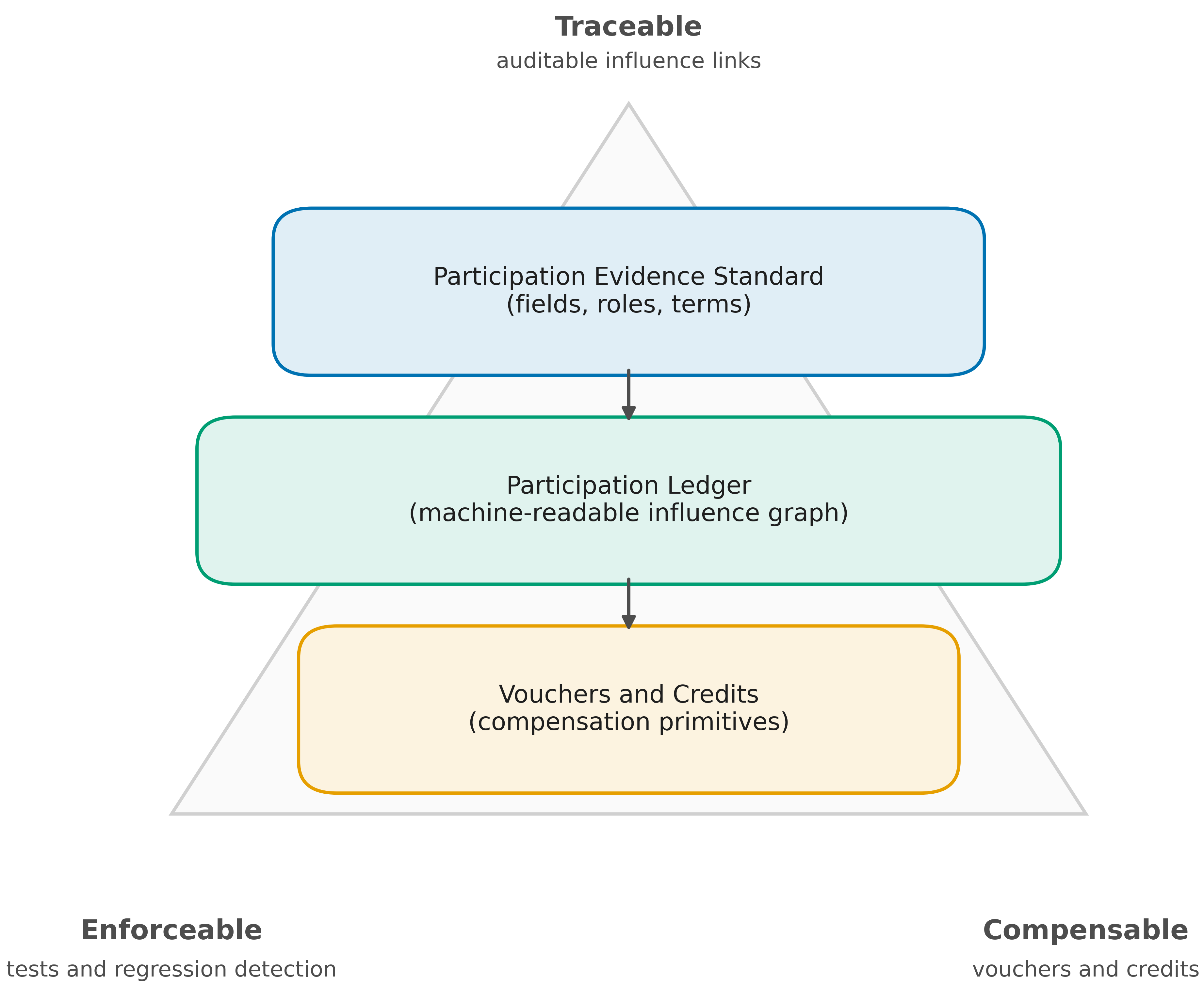}
  \caption{Design alignment: the Participation Evidence Standard makes participatory influence traceable; the Participation Ledger makes checks enforceable via replayable tests; vouchers and credits support compensable participation within an adoption boundary.}
  \Description{A triangle diagram whose corners are labeled TRACEABLE, ENFORCEABLE, and COMPENSABLE. Inside the triangle are three stacked boxes: Participation Evidence Standard, Participation Ledger, and Vouchers plus Credits, with downward arrows showing how evidence fields feed the ledger and enable vouchers and credits.}
  \label{fig:design_alignment}
\end{figure}

Figure~\ref{fig:design_alignment} summarizes the design alignment: traceability is provided by auditable influence links and evidence fields; enforceability is provided by replayable tests and regression detection tied to deployment checkpoints; compensability is provided by voucher and credit primitives that can be adopted within a defined organizational boundary.

The design is grounded in four documented cases at the intersection of AI, HCI, urban planning, and public-sector governance: AIAI/Mid-Space/LIVS \citep{Mushkani2025LIVS,Nayak2024MidSpace,MilaAIAIPage}, EVADIA+ \citep{EVADIAProjectPage}, AI-EDI-Space and Street Review \citep{Mushkani2024AIEDI,Mushkani2026StreetReview}, and WeDesign+ \citep{Mushkani2025WeDesign}.

The Participation Ledger is differentiated from existing documentation and provenance approaches. Datasheets and model cards summarize dataset and model properties, but they do not represent fine-grained, auditable links from particular participatory artifacts to subsequent changes or to replayable tests that can be run on later versions \citep{Gebru2021Datasheets,Mitchell2019ModelCards}. Decision provenance characterizes information flows and accountability across systems, but it does not, by itself, specify participation evidence fields, boundary-scoped rights, or crediting rules \citep{Singh2018DecisionProvenance}. Transparency logs and supply-chain attestations support integrity of artifact histories, but they do not provide semantics for participatory influence, nor do they connect influence claims to tests of change \citep{TorresArias2019InToto,Sigstore2023,SLSA2024}. Participatory AI and redress frameworks motivate inclusive processes and procedural safeguards, but they typically remain at the level of methods and institutional commitments rather than portable, machine-readable artifacts that can be audited across vendors and model updates \citep{Birhane2022PowerToPeople,Cobbe2021ReviewableADM,Reisman2018AIA}.

\section{Conceptual framing: participation as traceable influence}

\subsection{What counts as participation in this work}

Participation is a contested term with long-standing distinctions between symbolic inclusion and shared power \citep{Arnstein1969,CookeKothari2001}. In participatory design traditions, participation is not only a method but also an orientation toward shifting expertise and decision authority \citep{CostanzaChock2020,McKercher2020}. In planning and public-sector contexts, participation often occurs through consultations, deliberations, and co-design sessions, where the legitimacy of outcomes depends on how knowledge is incorporated and how commitments are upheld \citep{Arnstein1969}.

This work operationalizes participation as \emph{documented and actionable influence} that can be linked to changes in an AI system or its governance boundary. Participation includes, but is not limited to, prompt writing and refinement for generative systems, annotation and preference labeling, deliberation rationales recorded during workshops, incident and harm reports during deployment, and stewardship actions that condition capabilities. Participation also includes contributions from experts and intermediaries when they act as community stewards or facilitators, but the ledger requires that such roles be made explicit to avoid conflating proxy participation with direct involvement of affected groups \citep{Sloane2020ParticipationNotFix,Birhane2022PowerToPeople}. In other words, participation is treated as a set of contributions embedded in a sociotechnical process rather than a single event.

\subsection{Why traceability, enforceability, and compensation are linked}

Traceability, enforceability, and compensation are treated as linked requirements because each depends on the others in common governance scenarios. Traceability concerns whether a contribution can be connected to specific decisions, artifacts, and behavioral changes across model versions and organizational boundaries. Without traceability, communities and auditors cannot reliably assess whether participation had material effect, and organizations cannot reliably attribute responsibility for design choices or respond to contestation \citep{Singh2018DecisionProvenance,Gebru2021Datasheets}. Traceability also matters for long-lived systems, where the effects of earlier decisions persist in datasets, fine-tuning runs, and evaluation suites \citep{Mitchell2019ModelCards,NIST_AIRMF_2023}.

Enforceability concerns whether participation produces actionable rights, such as the ability to pause a capability, condition deployment on passing specific tests, or initiate remediation workflows. This is aligned with accountability frameworks emphasizing reviewability, contestability, and procedural safeguards, especially in public-sector contexts \citep{Cobbe2021ReviewableADM,Reisman2018AIA,Canada_Directive_ADM_2019}. Without enforceability, participation can be reduced to consultation, where community input is recorded but cannot shape outcomes when tensions arise, including when harms are reported or when vendors change models \citep{CookeKothari2001,Sloane2020ParticipationNotFix}.

Compensation concerns whether participation is recognized as labor and whether its value is sustained beyond a single engagement. Many AI systems depend on continuous work such as labeling, evaluation, monitoring, and incident reporting, and this work is often undervalued or invisibilized \citep{GraySuri2019GhostWork,10.1145/3706598.3713112}. In participatory settings, compensation can also be a condition for equitable engagement, especially when participation entails time, expertise, emotional labor, or exposure to harmful content \citep{CostanzaChock2020}. Without compensation mechanisms, participation becomes difficult to sustain and may exacerbate inequities by privileging those with greater time and resources.

These requirements form a cycle. Traceability enables enforceability by making commitments and influence paths auditable. Traceability also enables compensation by attributing ongoing value to specific contributions. Enforceability can create incentives for organizations to maintain traceability, because rights become operational only when linked to verifiable evidence and tests. Compensation can also motivate sustained oversight and reporting, but only if crediting can be audited and is not solely discretionary \citep{10.1145/3706598.3713112,GraySuri2019GhostWork}. The Participation Ledger is designed to bind these requirements within a single lifecycle-integrated artifact.

\section{Related work}

\subsection{Participation in AI development and governance}

Work on participatory AI argues that affected communities should shape problem framing, data practices, evaluation, and deployment decisions, but also notes risks of tokenism, burdens, and power asymmetries \citep{Birhane2022PowerToPeople,Sloane2020ParticipationNotFix}. In HCI and civic technology, crowdsourcing and participatory methods have been studied as ways to broaden engagement in governance, yet these methods raise questions about representativeness, legitimacy, and who benefits \citep{Brabham2009Crowdsourcing,AitamurtoChen2017CrowdsourcingPolicy}. In urban planning, participation has a long history, and the literature emphasizes that meaningful participation involves authority, transparency of decision pathways, and accountability for commitments \citep{Arnstein1969,CookeKothari2001}.

In responsible AI, participation is often recommended but inconsistently operationalized. Governance frameworks emphasize documentation, risk management, impact assessments, and post-deployment monitoring, but they rarely provide concrete mechanisms for binding community input to system changes or for attributing ongoing value to participatory artifacts \citep{NIST_AIRMF_2023,EU_AI_Act_2024,Reisman2018AIA}. Recent work on pluralistic alignment and value-sensitive system design highlights that multiple communities can hold divergent values, which complicates the idea of a single objective function and makes evidence about whose values were used and how they were aggregated particularly important \citep{Mushkani2025LIVS,Sorensen2024,Wallach2010MoralMachines}. The Participation Ledger contributes to this space by proposing a concrete representation for participation and its influence, designed to operate across vendors and over time.

\subsection{Participatory evaluation and tests}

Evaluation in AI has expanded beyond accuracy to include values such as fairness, safety, and context-specific performance, often using benchmarks, audits, and participatory assessments \citep{Raji2020AuditGap,Sandvig2014AuditingAlgorithms}. In participatory settings, evaluation is frequently conducted through workshops, user studies, and deliberations, which can surface harms not captured by standard metrics. However, participatory findings are often encoded as qualitative reports rather than as persistent tests that can be replayed on future versions \citep{Sloane2020ParticipationNotFix,Mushkani2024AIEDI,gerdes_participatory_2022}. In software engineering, regression tests and continuous integration practices provide mechanisms to prevent reintroduction of known defects. Analogous approaches have been proposed for AI evaluation pipelines, including model cards, system cards, and incident reporting practices \citep{Mitchell2019ModelCards,NIST_AIRMF_2023}. The Participation Ledger uses the concept of \emph{tests of change} as replayable, versioned artifacts that connect community concerns to concrete behavioral checks across updates.

\subsection{Documentation, provenance, and auditability}

Dataset and model documentation practices provide structured ways to record intended use, data collection context, limitations, and ethical considerations \citep{Gebru2021Datasheets,Mitchell2019ModelCards,BenderFriedman2018DataStatements}. These practices support accountability but do not, by themselves, ensure traceability from participation artifacts to specific changes. Provenance research, including decision provenance, argues that transparent information flows can support accountability in complex systems, particularly across organizational boundaries \citep{Singh2018DecisionProvenance}. Standardized provenance models, such as the W3C PROV family, and linked-data approaches, such as JSON-LD, enable interoperable representation of entities, activities, and agents \citep{W3C_PROV_2013,JSONLD_2020}. In security and software supply-chain contexts, transparency logs and cryptographic attestations help ensure integrity of build artifacts and changes, providing a basis for independent auditing \citep{TorresArias2019InToto,SLSA2024,Sigstore2023}. The Participation Ledger draws from these traditions to propose an append-only, auditable influence graph for participatory AI artifacts.

\subsection{Redress, contestability, and procedural rights}

Redress and contestability are recurring concerns in algorithmic governance, especially for systems affecting rights and access to services \citep{Cobbe2021ReviewableADM,Reisman2018AIA,EU_AI_Act_2024}. Impact assessment frameworks propose structured processes for identifying risks, engaging stakeholders, and documenting mitigation, but the operationalization of community rights often remains external to the technical system \citep{Reisman2018AIA,Canada_Directive_ADM_2019}. In public-sector procurement, accountability frequently depends on contractual obligations, audit rights, and post-deployment monitoring requirements. Yet, vendors may change models or service components over time, which can undermine earlier commitments unless they are anchored to verifiable evidence and enforcement mechanisms \citep{NIST_AIRMF_2023}. The Participation Ledger introduces Capability Vouchers as a rights mechanism that can be enforced within an adoption boundary by linking governance actions to verifiable artifacts and tests.

\subsection{Incentives, compensation, and the political economy of AI participation}

Compensation and incentives relate to the broader political economy of data and AI labor. Research on crowd work and data labor documents how annotation and evaluation tasks can be invisibilized, underpaid, and unevenly distributed \citep{GraySuri2019GhostWork,10.1145/3706598.3713112}. In participatory and community-based research, compensation is also an ethical consideration tied to equity, power, and the sustainability of partnerships \citep{Israel1998CBPR,CostanzaChock2020}. In AI governance, incentives may be necessary for sustained oversight, including community-driven monitoring and reporting of harms. However, compensation mechanisms can also introduce risks of coercion, gaming, and inequity if not carefully governed. The Participation Ledger specifies Participation Credits as a mechanism for attribution and compensation that is linked to measurable ongoing value, such as detection of regressions, while emphasizing explicit governance constraints and auditability.

\section{Field cases and requirement derivation}
\label{sec:cases}

\subsection{Method: cross-case synthesis across participatory urban AI projects}

We ground the Participation Ledger specification in a document-based cross-case analysis of four participatory urban AI projects. The aim is not to evaluate deployments, but to extract requirements from how participation is documented across lifecycle stages and across organizational boundaries. The cases were selected because they span participation modalities (workshops, interviews, preference labeling, and consultation processes) and artifact types (datasets, prompt repositories, evaluation protocols, and decision-support outputs), while remaining within a shared governance context of urban planning and public space.

\paragraph{Data sources.} For each case, we collected the project materials that describe participation and the resulting AI artifacts. For AIAI/Mid-Space/LIVS, the corpus comprises the Mid-Space documentation and project site, the LIVS paper, and the AIAI program page \citep{Nayak2024MidSpace,Mushkani2025LIVS,MilaAIAIPage}. For EVADIA+, the corpus is the public project documentation \citep{EVADIAProjectPage}. For AI-EDI-Space and Street Review, the corpus comprises the published descriptions of participation, data collection, and modeling pipelines \citep{Mushkani2024AIEDI,Mushkani2026StreetReview}. For WeDesign+, the corpus comprises the WeDesign+ paper and the project pilot notes used in the study \citep{Mushkani2025WeDesign}. Across cases, we extracted any reported quantities (for example, numbers of prompts, images, annotations, interviews) and any explicit statements about consent, privacy, compensation, reuse, and traceability mechanisms.

\paragraph{Extraction and synthesis.} We performed structured extraction into a case matrix with rows corresponding to cases and columns corresponding to (i) participation artifacts, (ii) lifecycle stages where artifacts appear, (iii) reported change and evaluation practices, and (iv) governance-relevant evidence elements. We then synthesized recurring requirements by comparing where cases rely on narrative reporting, where links between contributions and changes remain implicit, and where governance commitments are stated without machine-readable counterparts.

\paragraph{Coding for evidence coverage.} Table~\ref{tab:evidence} reports coverage of five evidence elements that the Participation Evidence Standard requires: recruitment pathway, roles and intermediaries, consent and privacy scope, compensation terms, and explicit influence links. For each element, we used conservative coding rules. \emph{Reported} means that the case materials explicitly specify operational details that could be transcribed into a ledger entry (for example, a stated recruitment channel, a role description that distinguishes contributors from facilitators, a described consent scope or privacy constraint, a stated compensation model or amount, or an explicit description of how participatory artifacts are used to trigger or justify specific changes). \emph{Partial} means that the element is mentioned at a high level but omits operational terms needed for audit or reuse (for example, consent is referenced without reuse boundaries, compensation is mentioned without conditions, or influence is described only as a general pipeline without contribution-to-change links). \emph{Not specified} means that we did not find an explicit statement in the corpus. When information was ambiguous or conditional, we coded it as not specified rather than inferring practices.

Table~\ref{tab:cases} summarizes the cases and the participation artifacts that motivate ledger mechanisms.
Extended narratives (and a visualization of reported artifact volumes) are provided in Appendix~\ref{app:case_narratives}.

\begin{table*}[t]
\caption{Field cases grounding the Participation Ledger design (from public project materials; see Section~\ref{sec:cases}). Reported quantities are descriptive indicators; not all cases report all quantities.}
\label{tab:cases}
\scriptsize
\setlength{\tabcolsep}{4pt}
\renewcommand{\arraystretch}{1.15}
\begin{tabularx}{\textwidth}{@{}lX X X@{}}
\toprule
Case & Context and stakeholders & Participation artifacts & Reported scale indicators \\
\midrule
AIAI / Mid-Space / LIVS & Inclusive visualization of public spaces in Montreal with community organizations, facilitators, and researchers. & Workshops and interviews; prompt writing; multi-criteria preference labeling; identity markers (LIVS); iteration on annotation tooling. & Mid-Space: 3,350 prompts; 16,694 images; >42k annotations. LIVS: 13,462 images plus preference labels with splits. \\
EVADIA+ & Civic participation and planning decision support that translates consultation inputs into indicators and recommendations. & Consultation-derived qualitative data; co-constructed indicators; process documentation intended for decision-makers. & Public documentation emphasizes goals and process; artifact volumes are not consistently reported. \\
AI-EDI-Space / Street Review & Participatory evaluation of public spaces and streetscapes with image-based ratings and modeling. & Interviews and discussion panels; image ratings or pairwise comparisons; criteria refinement; model validation against human judgments. & AI-EDI-Space: 10,000 photos; 7,833 labeled points; 30 participants. Street Review: applied to >45k street-view images; training labels from 15,000 images across 60 locations. \\
WeDesign+ & Participatory text-to-image co-creation for urban design consultations with mixed stakeholder workshops and expert interviews. & Prompt writing and iterative edits; negotiation rationales; feature and safety requests for participatory tooling. & Workshop with 29 participants; 440 prompts; qualitative coding with agreement procedure. \\
\bottomrule
\end{tabularx}
\end{table*}

\subsection{Cross-case requirements}
\label{sec:case_requirements}

Across cases, participation is documented primarily in narrative form and rarely travels as structured, auditable metadata across model versions, vendor boundaries, and deployments. We use the cases to derive a minimal set of requirements that the ledger must satisfy (extended case narratives are in Appendix~\ref{app:case_narratives}):

\begin{itemize}
  \item \textbf{Portable participation evidence:} record recruitment pathways, role boundaries (including intermediaries), consent/privacy scope, compensation terms, and reuse/retention constraints as machine-readable fields.
  \item \textbf{Influence links that survive iteration:} assign stable identifiers to prompts, labels, criteria, workshop rationales, and incident reports, and explicitly link them to concrete changes (UI, data, model, policy) rather than relying on post-hoc narrative.
  \item \textbf{Replayable participatory evaluation:} translate salient community concerns into \emph{tests of change} that can be rerun on later versions to detect regressions and support contestation.
  \item \textbf{Boundary-scoped enforcement:} represent rights that an adopting organization can actually enforce (e.g., release gates, scoped capability restrictions, pause requests) rather than assuming global control over a model supply chain.
  \item \textbf{Sustained credit and compensation:} make ongoing maintenance value legible (e.g., when contributed tests prevent regressions or trigger remediation), enabling auditable attribution and compensation policies.
\end{itemize}

\section{The Participation Ledger}
\label{sec:ledger}

\subsection{Overview: an auditable influence graph}

The Participation Ledger is designed as a machine-readable influence graph that records participation artifacts and links them to decisions, changes, and verifiable tests. The ledger adopts provenance concepts in which agents perform activities that generate entities, and where edges represent influence relationships \citep{W3C_PROV_2013,Singh2018DecisionProvenance}. The ledger is intended to be auditable in the sense that independent reviewers can trace a chain from a contribution to a system modification and to a test that demonstrates behavioral change. The ledger is also intended to be portable across organizations by using a structured schema that can be serialized as JSON-LD, enabling integration with existing documentation practices and toolchains \citep{JSONLD_2020}.

\begin{figure}[t]
  \centering
  \includegraphics[width=0.9\linewidth]{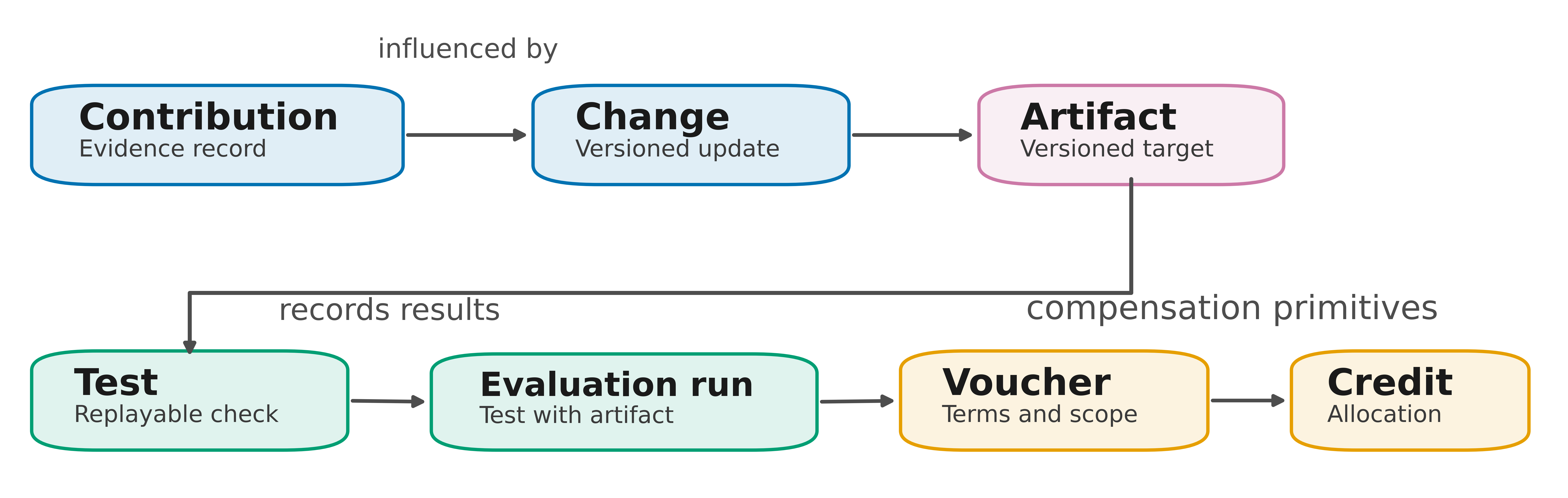}
  \caption{Core Participation Ledger schema. Contributions influence changes; changes produce versioned artifacts and require replayable tests; evaluation runs bind tests to specific artifact versions; vouchers and credits represent governance and compensation primitives within an adoption boundary.}
  \Description{A diagram of the core Participation Ledger entity types and links. A Contribution influences a Change, which produces an Artifact and requires a Test. The Test and Artifact feed an Evaluation run. The Evaluation run earns a Credit, and a Voucher allocates the Credit.}
  \label{fig:ledger}
\end{figure}

Figure~\ref{fig:ledger} sketches the ledger as a minimal influence graph: participation artifacts (\texttt{Contribution}) are linked to system modifications (\texttt{Change}); each change produces versioned targets (\texttt{Artifact}) and is paired with one or more replayable \emph{tests of change} (\texttt{Test}). Executing a test on a specific artifact version yields an \texttt{EvaluationRun} with recorded inputs, procedure, and outcome. The same structure supports governance and incentives within an adoption boundary: Capability Vouchers record authorized requests to pause or condition a capability, and Participation Credits record attribution events when contributed tests or artifacts continue to provide measurable maintenance value.

In this schema, participation artifacts such as prompts, annotations, deliberation rationales, and incident reports are recorded as \texttt{Contribution} nodes. System modifications—dataset updates, prompt library revisions, adapter fine-tuning runs, or policy/guardrail changes—are recorded as \texttt{Change} nodes and must cite the motivating contributions. Versioned targets (models, datasets, prompt libraries, policies) are recorded as \texttt{Artifact} nodes. \texttt{Test} nodes capture replayable checks derived from participatory concerns; each execution is recorded as an \texttt{EvaluationRun} bound to a specific artifact version and decision. Vouchers and credits sit alongside this traceability substrate to support boundary-scoped rights and auditable attribution events.

\subsection{Participation Evidence Standard}

The Participation Evidence Standard specifies the minimum evidence required for a ledger entry to be considered complete for audit and governance purposes. The standard is intentionally conservative: it requires fields that enable informed reuse decisions and accountability, even if particular projects cannot initially populate all fields. The standard is motivated by recurring documentation practices in responsible AI (e.g., datasheets and model cards) but extends them to participation-specific concerns such as recruitment gaps, consent scope, and compensation terms \citep{Gebru2021Datasheets,Mitchell2019ModelCards}.

In the ledger, each \texttt{Contribution} entry includes: a stable identifier; the modality and artifact type (e.g., prompt, preference label, interview excerpt, deliberation rationale, incident report); the date and context of creation; a description of the role of the contributor (e.g., resident participant, community steward, facilitator, researcher); consent status and scope; privacy constraints and retention expectations; compensation terms and any restrictions on downstream use; and, where applicable, representational metadata that supports assessing recruitment coverage. Representational metadata is treated as optional and sensitive: the standard permits pseudonymous or aggregated identity markers and requires explicit consent for any demographic or identity-related fields \citep{Crenshaw1989,Gebru2021Datasheets,Mushkani2025LIVS}.

The standard also requires fields that support governance and procurement. Each contribution includes a declared intended use (e.g., evaluation-only, training, documentation), and a set of enforceable constraints expressed as policy tags that can be referenced in procurement clauses or internal governance policies. For example, a community organization may permit use of prompts for evaluation but not for training, or may permit training use only if the resulting model is restricted to non-commercial deployment in specific contexts. The standard does not assume that these constraints are automatically enforceable across all institutions; rather, it requires that constraints be recorded in a machine-readable form so that enforcement can be audited and contractualized \citep{Canada_Directive_ADM_2019,EU_AI_Act_2024}.

\subsection{Influence tracing}

Influence tracing is the mechanism by which the ledger connects contributions to changes and changes to system behavior. The ledger models influence as a directed graph with explicit edges such as \texttt{influencedBy} and \texttt{influences}. A change entry must include references to the contributions and decisions that motivated it, and must declare the artifacts it modifies (e.g., dataset version, prompt repository, model adapter, guardrail rules). Influence tracing does not claim to capture causal attribution in the philosophical sense. Instead, it operationalizes traceability as a record of what maintainers claim to have changed in response to what participation artifacts, with integrity mechanisms intended to make after-the-fact rewriting detectable \citep{Singh2018DecisionProvenance,TorresArias2019InToto}.

In the four cases, influence paths often exist informally. In AIAI, prompts and preference labels influence model fine-tuning and evaluation, and workshop feedback influences annotation tooling changes \citep{Mushkani2025LIVS,Nayak2024MidSpace}. In WeDesign+, participant prompt edits and discussions influence prompt libraries and feature requests \citep{Mushkani2025WeDesign}. In AI-EDI and Street Review, interviews and ratings influence criteria definitions and training labels, which influence model outputs used for planning analysis \citep{Mushkani2024AIEDI,Mushkani2026StreetReview}. Influence tracing formalizes these connections by requiring each change to include pointers to the specific evidence artifacts and by requiring tests that demonstrate behavioral implications.

\subsection{Tests of change}

A test of change is a replayable artifact that demonstrates how a system behaves before and after a change, for a specific concern or requirement. Tests can be qualitative (e.g., a scenario prompt and a human rating protocol), quantitative (e.g., a metric threshold on a fixed evaluation set), or hybrid (e.g., a rubric scored by human raters with aggregation rules). A test includes an input specification, an expected behavior specification, and a measurement procedure that can be repeated across versions.

In a generative model workflow, a test might specify a prompt and an evaluation rubric related to accessibility features, such as ramps, wide pathways, or signage, reflecting criteria derived in Mid-Space and LIVS \citep{Nayak2024MidSpace,Mushkani2025LIVS}. In WeDesign+, a test might specify that a prompt template used in consultation should not produce outputs that violate a safety constraint or that systematically omit certain accessibility features, with evaluation based on a participant-defined rubric \citep{Mushkani2025WeDesign}. In AI-EDI and Street Review, a test might specify that predicted scores for a set of images should remain within defined bounds relative to participant distributions, or that model calibration should not regress on specific neighborhoods or demographic groups, within the constraints of available data \citep{Mushkani2024AIEDI,Mushkani2026StreetReview}.

Tests of change are intended to be linked to the contributions that motivated them. For example, an incident report about repeated failure to represent mobility aids in generated images can be converted into a test that is automatically rerun on future model versions. This supports a notion of participatory regression testing, where community concerns are encoded as durable evaluation artifacts rather than as one-time feedback.

\paragraph{Illustrative lifecycle.} In a WeDesign+-style consultation, participants may surface an accessibility concern (e.g., generated images omit continuous wheelchair-accessible paths). The ledger records the prompt and rationale as a \texttt{Contribution} with consent and compensation terms; maintainers record the remediation \texttt{Change} and bind it to a \texttt{Test} of change (a prompt set and rubric); and \texttt{EvaluationRun} entries record outcomes on each model version. If a later update fails the same test, a steward can issue a Capability Voucher to pause the capability within the adoption boundary until remediation and re-evaluation. Appendix~\ref{app:ledger_workflow} provides an end-to-end worked example and the evaluation-harness workflow.

\subsection{Capability Vouchers}

Capability Vouchers are a rights mechanism specified within the ledger. A voucher is a machine-readable authorization or constraint that allows an authorized steward entity to request, pause, or condition a capability within a defined boundary. The boundary is critical: the ledger cannot unilaterally enforce rights across organizations, but it can provide a verifiable record of requests and conditions that can be enforced within an adopting organization's deployment controls or through contractual obligations with vendors \citep{EU_AI_Act_2024,Reisman2018AIA}.

A capability is defined as a deployment-relevant function, such as the ability to generate images of a public space design for consultation, to provide a neighborhood-level inclusivity score map, or to summarize consultation transcripts into recommendations. A voucher can specify constraints such as restricting capability use to a given context, requiring that specific tests pass before activation, limiting output resolution or retention, or requiring human-in-the-loop review for certain outputs. The voucher mechanism is designed to be compatible with credential and policy approaches, such as verifiable credentials for authorization and policy tags for enforcement \citep{W3C_VC_2019}.

Capability Vouchers respond to governance needs surfaced in EVADIA+ and the evaluation-focused cases. In planning contexts, a community might request that a decision-support output not be used for high-stakes decisions until specific participatory evaluation tests are satisfied, or that outputs be accompanied by uncertainty and limitation disclosures. Vouchers provide a structured way to record such requests and their resolution, supporting accountability and reviewability \citep{Cobbe2021ReviewableADM}.

\subsection{Participation Credits}

Participation Credits are an incentive and compensation mechanism specified within the ledger. Credits are accrued when participation artifacts provide ongoing value, such as when a community-contributed test detects a regression during a model update, or when a prompt template is reused as part of a standardized evaluation suite. Credits are not defined as a universal currency; instead, the ledger represents them as auditable units of attribution that can be mapped to compensation arrangements within an adopting organization or contract. This design reflects constraints identified in the political economy of data labor: compensation mechanisms must be explicit about who benefits, under what terms, and how ongoing value is measured \citep{GraySuri2019GhostWork,10.1145/3706598.3713112}.

The credit mechanism includes two design principles. First, crediting is linked to measurable events recorded in the ledger, such as a test failure detection, a remediation completion, or a scheduled evaluation run that depends on contributed artifacts. Second, crediting is separable from identity disclosure: contributors can be credited via pseudonymous identifiers or steward organizations, reducing the risk that crediting becomes a vector for surveillance \citep{CostanzaChock2020}. Credits can also be aggregated at the group level when participation occurs collectively, as in workshop settings, aligning with how co-creation often operates in WeDesign+ and AIAI \citep{Mushkani2025WeDesign,Mushkani2025LIVS}.

\section{Implementation and integration}
\label{sec:implementation}

The Participation Ledger is designed to attach to artifacts and control points that already exist in AI development and public-sector adoption workflows: version control commits, dataset releases, model registries, evaluation dashboards, and deployment configuration. In practice, maintainers can emit ledger entries when prompts, datasets, models, or policies change, while stewards manage evidence fields (recruitment, consent scope, compensation) and voucher authorizations.

\subsection{Representation and validation}

We represent ledger exports as JSON-LD aligned with W3C PROV semantics to support portable graph data and interoperable tooling \citep{JSONLD_2020,W3C_PROV_2013}. The core classes are \texttt{Contribution}, \texttt{Change}, \texttt{Artifact}, \texttt{Test}, \texttt{EvaluationRun}, \texttt{Voucher}, and \texttt{Credit}. As a minimal feasibility check, we implemented a reference JSON Schema for these classes and validated representative entries to ensure that Participation Evidence Standard fields and influence links can be captured and machine-checked. Appendix~\ref{app:schema} provides a compact schema excerpt and example serialization.

\subsection{Auditability within an adoption boundary}

For auditability, the ledger assumes an append-only log with content hashes (and optional signatures) so that rewriting or deletion becomes detectable, drawing on transparency logging and software supply chain integrity practices \citep{TorresArias2019InToto,Sigstore2023,SLSA2024}. Enforcement of vouchers and credit policies is boundary-scoped: it relies on contractual levers (e.g., procurement acceptance criteria) and technical levers (e.g., CI/CD release gates and configuration management) that the adopting organization controls. Appendix~\ref{app:auditability} details integrity mechanisms, a threat model, and example audit queries.

\section{Evaluation}
\label{sec:evaluation}

This paper specifies a ledger and demonstrates its need and feasibility using available artifacts, without claiming downstream deployment impact. We therefore evaluate (i) how well the participation evidence fields we require are currently documented in exemplary participatory urban AI projects, and (ii) whether the specification can be represented as a machine-validated schema (Section~\ref{sec:implementation}).

\subsection{Evidence gaps surfaced by the cases}

We coded publicly available documentation for the four cases to assess whether key Participation Evidence Standard elements are explicitly specified. The goal is not to grade the projects, but to surface which elements are routinely underspecified in materials that otherwise report rich participatory methods.

\begin{table}[t]
\caption{Document-based coverage of Participation Evidence Standard elements across cases. Values are coded from the case documents cited in Section~\ref{sec:cases}. ``Partial'' indicates that an element is mentioned but lacks operational terms needed for audit or reuse.}
\label{tab:evidence}
\small
\begin{tabularx}{\linewidth}{@{}l>{\raggedright\arraybackslash}X>{\raggedright\arraybackslash}X>{\raggedright\arraybackslash}X>{\raggedright\arraybackslash}X>{\raggedright\arraybackslash}X@{}}
\toprule
Case & Recruitment pathway & Roles and intermediaries & Consent and privacy scope & Compensation terms & Explicit influence links \\
\midrule
AIAI / Mid-Space / LIVS & Reported & Partial & Partial & Not specified & Partial \\
EVADIA+ & Partial & Partial & Partial & Not specified & Not specified \\
AI-EDI-Space / Street Review & Reported & Partial & Partial & Not specified & Partial \\
WeDesign+ & Reported & Partial & Partial & Reported & Partial \\
\bottomrule
\end{tabularx}
\end{table}

Across cases, recruitment pathways are sometimes described, but consent scope, compensation terms, and explicit influence links from participation artifacts to concrete system changes are often absent or too high-level to support audit across versions and vendors. This motivates making these fields explicit and machine-readable in the Participation Evidence Standard, and binding them to influence tracing and tests of change.

\subsection{Stakeholder and deployment evaluation plan}
\label{sec:stakeholder_validation}

Document analysis cannot tell us which evidence fields and enforcement levers are \emph{practically} enforceable within adopting institutions. We therefore provide an interview protocol (Appendix~\ref{app:interview_guide}) to validate feasibility across procurement staff, community stewards, and ML maintainers. For deployments, we propose measuring linkage completeness and audit success (can an independent reviewer reconstruct an influence path and verify a test within a bounded time), boundary-scoped enforcement (whether vouchers condition releases), and crediting behavior under explicit policies, including adversarial scenarios such as selective omission or credit gaming \citep{Sandvig2014AuditingAlgorithms,10.1145/3706598.3713112}.

\section{Governance and procurement implications}

\subsection{Accountability across vendors and institutions}

Public-sector AI systems often involve multiple vendors and changing components over time. Accountability failures can arise when a system changes without corresponding updates to documentation, evaluation, or commitments made during consultations \citep{Singh2018DecisionProvenance,NIST_AIRMF_2023}. The Participation Ledger is designed to support cross-boundary accountability by providing portable evidence and by attaching governance constraints to artifacts. In procurement contexts, a ledger can function as a condition of purchase by specifying what evidence must be delivered with a system, how evaluation tests are maintained, and how governance actors can request remediation.

The ledger does not guarantee accountability. It is a representational infrastructure that can support accountability if paired with institutional commitments, audit rights, and enforcement mechanisms. For example, procurement clauses can require that a vendor provide ledger exports for each release, that evaluation runs be reproducible, and that incident response includes ledger-updated tests. These clauses are proposed templates rather than claims about existing contractual adoption in the cases described here. Appendix~\ref{app:procurement} provides a reference clause template for discussion.

\subsection{Redress workflows}

The ledger supports a structured redress workflow that links a reported harm to verification and remediation. A redress workflow begins with an incident report recorded as a contribution, with consent and privacy terms. The incident is triaged into a test of change that captures the concern in a replayable form. A remediation change is then recorded, linked to the test, and validated through an evaluation run. The ledger can record communications and decisions associated with the redress process, supporting reviewability and accountability \citep{Cobbe2021ReviewableADM,Reisman2018AIA}. Participation Credits can be triggered when a contributed test detects a regression or supports maintenance over time, making continued oversight legible and potentially compensable.

\section{Ethics and limitations}

\subsection{Consent, privacy, and data protection}

Ledgering participation raises privacy risks because it links people to artifacts and potentially to sensitive identity markers. The Participation Evidence Standard therefore treats consent scope as a required field and supports restricted and withdrawn consent states. The ledger design encourages data minimization: storing references and hashes rather than raw content when possible, supporting pseudonymous identifiers, and separating identity markers from publicly shared exports. These choices align with standard privacy and documentation principles in responsible AI and provenance systems \citep{Gebru2021Datasheets,W3C_VC_2019}. However, privacy trade-offs remain. For example, auditability can conflict with the right to withdraw or with safety concerns if a ledger makes participation traceable in ways that expose community members. Implementations must define governance processes for redaction, aggregation, and controlled access, and must treat participation data as sensitive.

\subsection{Power, tokenism, and representational harms}

Participation mechanisms can reproduce power asymmetries when decision authority remains centralized, when recruitment is unrepresentative, or when participation is used to legitimize predetermined outcomes \citep{Arnstein1969,CookeKothari2001,Sloane2020ParticipationNotFix}. The ledger does not resolve these structural issues, but it can make them more legible by requiring recruitment and representational reporting fields and by recording whether and how participation influenced changes. This legibility can support accountability, but it can also create incentives for performative compliance if organizations focus on filling fields rather than shifting power. Governance processes must therefore include independent oversight and qualitative review of participation quality, not only quantitative ledger completeness.

Representational harms can occur when participatory artifacts are used to essentialize groups or when identity markers are interpreted without context \citep{Crenshaw1989,Benjamin2019}. LIVS introduces identity markers to support pluralistic analysis, but such markers are sensitive and require careful consent and governance \citep{Mushkani2025LIVS}. The ledger treats identity markers as optional and emphasizes that any use must be justified, consented, and limited.

\subsection{Risks introduced by ledgering}

Ledgering introduces new risks beyond those present in typical participatory documentation. Surveillance risks arise if ledger entries are used to track individuals, to evaluate participation behavior, or to penalize dissent. Coercion risks arise if compensation mechanisms pressure participation or if vouchers and credits become tied to gatekeeping within communities. Gaming risks arise if actors attempt to generate low-quality contributions to accrue credits or to strategically influence evaluation tests. The ledger also introduces bureaucratic risks: participation may become more burdensome if evidence requirements are imposed without support.

Mitigation strategies include governance by community stewards with transparent rules; independent audits; randomized review of credited events; caps and quality gates for credit accrual; and explicit separation between participation evidence and employment or surveillance systems. Implementations should also consider differential access, such as language accessibility and disability accommodations, which were salient in the urban consultation contexts of the field cases \citep{Mushkani2025WeDesign,EVADIAProjectPage}.

\subsection{Limitations and scope}

This paper specifies a Participation Ledger and grounds its requirements in four field cases, but it does not report a longitudinal deployment of the ledger in a production public-sector setting. Some claims are therefore design intents rather than measured impacts. The cross-case analysis is limited to what is reported in available documentation and cannot infer unreported consent or compensation practices. The ledger is also not a legal instrument; enforceability depends on organizational adoption, procurement contracts, and governance institutions. Finally, the ledger's emphasis on auditability can conflict with privacy and community safety, requiring careful governance choices that may differ across contexts.

\section{Conclusion}

This paper specified the Participation Ledger, a machine-readable influence graph for representing participation artifacts and linking them to changes, tests of change, evaluation runs, deployments, and governance actions. The specification integrates a Participation Evidence Standard for recording recruitment pathways, roles and intermediaries, consent and privacy scope, compensation terms, and reuse boundaries; an influence tracing mechanism that binds changes to replayable tests of change; and two ledger-representable governance primitives, Capability Vouchers and Participation Credits, that can be adopted within a defined boundary.

We grounded the specification in a document-based cross-case analysis of four participatory urban AI projects. The analysis is limited to what is reported in available documents, but it surfaces recurring gaps in how participation conditions and influence links are recorded. We provided a JSON-LD-oriented schema excerpt and reference templates and outlined an evaluation design for future deployments that measures linkage completeness, boundary-scoped enforcement behavior, and crediting under explicit governance policies.

Future work can evaluate the ledger in deployments where procurement, incident response, and oversight are active constraints, and can study how minimization, access control, and stewardship authority affect auditability and participant safety.

\section*{Generative AI Usage Statement}
Generative AI tools were used only for grammar and stylistic editing. All ideas, analyses, and written content are the authors’ own; no generative AI was used for data generation, analysis, or interpretation.

\bibliographystyle{unsrtnat}
\bibliography{references,software}

\clearpage
\appendix

\section{Supplementary ledger workflow example}
\label{app:ledger_workflow}

This appendix provides an end-to-end illustrative lifecycle for the ledger, complementing the main-text specification.

\subsection{Worked example: accessibility regression in a consultation workflow}

During a WeDesign+-style consultation workshop, participants note that prompts describing a waterfront promenade yield images that omit continuous wheelchair-accessible paths. A facilitator, acting as a steward, records a \texttt{Contribution} entry that includes the prompt text, a short rationale extracted from the workshop record, and the consent, privacy, and honoraria terms under which the artifact can be reused. Maintainers implement a \texttt{Change} that updates a shared prompt template and an output guardrail to require visible ramps and continuous paths. The change is linked to a \texttt{Test} of change consisting of a fixed prompt set and a participant-defined rubric for scoring accessibility features. An \texttt{EvaluationRun} executes the test on model version $v_1$ and records a pass decision, and the deployment boundary for the consultation tool records activation of the image-generation capability for $v_1$.

In a later update to $v_2$, the same test fails and the ledger records the regression in a new \texttt{EvaluationRun}. The steward issues a \texttt{Voucher} that pauses the capability within the adopting organization's deployment boundary until a remediation \texttt{Change} is recorded and the test passes again. After remediation and re-evaluation, a \texttt{Credit} entry attributes the regression detection to the original test contributors under the adopted credit policy, without requiring disclosure of personal identities.

\subsection{Participatory evaluation harness}

The participatory evaluation harness is a workflow that collects tests of change from the ledger and executes them on specified artifacts (e.g., model versions, prompt libraries) at defined checkpoints. Checkpoints can include pre-deployment gates, scheduled audits, and post-incident re-evaluations. The harness produces \texttt{EvaluationRun} entries that record inputs, outputs, metrics, and pass/fail decisions, linked to both the tests and the evaluated artifact versions.

The harness is designed to accommodate mixed evaluation methods. Some tests may require human evaluation, as in preference-based labeling of generative outputs \citep{Mushkani2025LIVS,Nayak2024MidSpace}. Others may rely on automated metrics, as in predictive model validation reported in Street Review \citep{Mushkani2026StreetReview}. The harness does not prescribe a single metric; instead, it records the measurement procedure, including who performed the evaluation, what consent and compensation terms apply to evaluators, and what aggregation method was used. This is intended to reduce ambiguity about responsibility and to support auditability when evaluation results are used to justify deployment decisions \citep{NIST_AIRMF_2023,Cobbe2021ReviewableADM}.

\section{Auditability mechanisms and threat model}
\label{app:auditability}

This appendix collects supplementary implementation details referenced in Section~\ref{sec:implementation}: field sensitivity and minimization guidance, example audit queries, integrity mechanisms, and threat-model assumptions.

\begin{figure}[t]
\centering
\small
\begin{tabularx}{\linewidth}{@{}>{\raggedright\arraybackslash}X>{\raggedright\arraybackslash}X>{\raggedright\arraybackslash}X@{}}
\toprule
Required fields (examples) & Optional fields (examples) & Sensitive fields (examples) \\
\midrule
Entry identifier and type; timestamps; actor role; artifact reference (URI or content hash); influence links; integrity hash and optional signature. & Steward co-signature; free-text justification; references to meeting agendas or minutes; retention schedule; redress contact; policy tags and procurement clause identifiers. & Direct identifiers; raw interview excerpts; demographic or identity markers; payment details; location traces; incident narratives beyond the agreed disclosure scope. \\
\bottomrule
\end{tabularx}
\caption{Ledger fields grouped by requirement and sensitivity. The field set supports minimization by allowing sensitive content to be stored out of band and referenced by hash under controlled access.}
\Description{A three-column table listing examples of required ledger fields, optional fields, and sensitive fields that should be access controlled.}
\label{fig:fields}
\end{figure}

\paragraph{Example audit query.} The ledger supports queries that combine participation evidence, influence links, and evaluation results. For example, an auditor can ask which participatory contributions created tests that later detected a regression after deployment within a specific adoption boundary.

\begin{lstlisting}
MATCH (c:Contribution)-[:MOTIVATES]->(t:Test)
MATCH (t)<-[:USES_TEST]-(r:EvaluationRun)-[:EVALUATES]->(a:Artifact)
MATCH (a)-[:DEPLOYED_AS]->(d:Deployment)
WHERE t.topic = "accessibility" AND r.decision = "fail"
  AND d.boundary = "consultation_workflow"
RETURN c.id, t.id, r.artifact_version, r.timestamp, d.id;
\end{lstlisting}

\subsection{Auditability mechanisms}

Auditability depends on both semantic traceability and integrity guarantees. The ledger's influence edges provide semantic traceability, but the system also needs mechanisms that make rewriting or selective deletion detectable. The ledger therefore assumes an append-only log structure, where each entry includes a content hash and, optionally, a hash pointer to the previous entry, forming a hash chain. This approach aligns with integrity practices in software supply chain security and transparency logging, where cryptographic attestation enables independent verification of artifact integrity \citep{TorresArias2019InToto,Sigstore2023,SLSA2024}. The ledger does not require a public blockchain; it can be implemented as a signed log within an organization, with periodic anchoring to a transparency service if appropriate.

Each ledger entry includes an \texttt{integrity} block, containing the hash of the canonical serialized entry and optional fields such as a previous-entry hash and a digital signature. Signatures support non-repudiation and role-based accountability when required, but they also raise privacy and coercion risks; therefore, the ledger allows pseudonymous signing and steward-level signing depending on governance choices \citep{W3C_VC_2019}. Auditability is further supported by strict versioning of referenced artifacts, such as dataset versions and model identifiers, to avoid ambiguity when systems evolve.

\subsection{Threat model and governance assumptions}

The Participation Ledger is a representational and audit artifact rather than a self-enforcing control plane. It is intended to make omissions, disputed claims, and policy violations more detectable within an adopting organization's boundary, and to make those issues more legible for external auditors when audit access is granted. This section states the governance assumptions needed for that goal and the failure modes the design explicitly anticipates.

\paragraph{Actors and permissions.} The ledger distinguishes at least five roles. Contributors create participation artifacts; stewards (often community organizations or delegated representatives) authorize collection, consent scope, and, where applicable, voucher actions; maintainers implement changes and record links from changes to contributions and tests; evaluators execute tests and record evaluation runs; and deployers operate the system within a defined adoption boundary. Auditors (internal procurement, oversight bodies, or independent reviewers) are not assumed to write entries, but are assumed to have read access to the relevant subset of ledger entries and referenced artifacts. The model assumes that stewards and maintainers can cryptographically sign entries, and that stewardship actions such as Capability Vouchers require steward authorization.

\paragraph{Adversaries and failure modes.} A primary risk is omission or selective linking, where a maintainer records a change but omits the contributions that motivated it, or links only a subset that is convenient. A second risk is post-hoc rewriting attempts, including editing past entries, deleting inconvenient entries, or rewriting artifact references after external scrutiny. A third risk is credit gaming, where actors create low-quality tests to accumulate Participation Credits, or reframe existing work as a novel contribution. A fourth risk is coercion and pressure, where participants are induced to consent to broad reuse or to accept compensation terms that they would not accept under low-power conditions. Finally, the ledger can be misused for surveillance if identity-linked participation artifacts are exposed beyond the contexts for which participants consented.

\paragraph{Mitigations consistent with the design.} Append-only logging with content hashes and optional hash chaining is intended to make rewriting or deletion detectable, especially when combined with signatures and, where feasible, periodic anchoring to a transparency service \citep{TorresArias2019InToto,Sigstore2023}. To reduce selective linking, the design allows stewards to co-sign changes or to issue a voucher that conditions deployment on the presence of specific links (for example, a change is not eligible for deployment without a linked test of change and an evaluation run). To reduce credit gaming, the credit policy can impose caps, quality gates, and persistence requirements (for example, credits accrue only for tests that remain in the evaluation set for a minimum number of releases and that materially affect pass/fail outcomes). To address coercion and privacy risks, the Participation Evidence Standard requires explicit consent scope and minimization, permits pseudonymous identifiers, and supports access controls so that sensitive fields can be withheld from broad distribution while still allowing audit under controlled access.

\paragraph{Limitations and trade-offs.} The ledger cannot ensure that all relevant participation is recorded; it can only make omissions easier to surface when governance actors have incentives and capacity to audit. Enforcement of vouchers and credit policies depends on the adoption boundary, including tooling integration and contractual authority. Minimization and access controls reduce surveillance risk but can also limit independent verification if auditors cannot access referenced artifacts. These trade-offs need to be resolved through governance choices that are explicitly recorded as part of the ledger configuration for a given deployment.

\section{Extended field case narratives}
\label{app:case_narratives}

This appendix provides extended descriptions of the four field cases referenced in Section~\ref{sec:cases}. The narratives are based on publicly available project materials and are used to illustrate the kinds of participation artifacts and documentation patterns that the ledger must accommodate.

\subsection{Case 1: AIAI, Mid-Space, and LIVS}

AIAI is framed as an applied research program using AI to support citizen engagement in co-creating urban spaces \citep{MilaAIAIPage}. Mid-Space and LIVS document a generative model workflow supported by community workshops and interviews, prompt creation, and multi-criteria labeling/preference collection (including identity markers in LIVS) \citep{Nayak2024MidSpace,Mushkani2025LIVS}.

The case illustrates participation spanning the full lifecycle (problem framing, data creation, evaluation, and iteration) while leaving influence links largely implicit: documentation describes how labels are used for training/evaluation, but does not provide an auditable chain from a specific piece of community feedback to a specific change in prompts, UI, or model behavior. The materials also do not describe boundary-scoped governance levers (e.g., a steward right to pause a capability) or durable compensation terms that travel with reused artifacts \citep{Mushkani2025LIVS}.

\subsection{Case 2: EVADIA+}

EVADIA+ positions AI as a mediator between citizen consultations and planning decision-making, emphasizing co-creation with citizens and experts and the translation of qualitative inputs into indicators and recommendations \citep{EVADIAProjectPage}. Here, participation artifacts are not only prompts/labels but also consultation-derived qualitative data, deliberation rationales, and indicators intended to travel to decision-makers.

The case motivates ledger support for cross-institution portability and procurement readiness: participation evidence must survive handoffs between organizations and vendors, and consent/privacy boundaries must be explicit for consultation data that may include sensitive information \citep{EVADIAProjectPage,Gebru2021Datasheets}.

\subsection{Case 3: AI-EDI-Space and Street Review}

AI-EDI-Space and Street Review use participatory methods (interviews, panels, image-based evaluations) to define criteria and to gather labels that train models for evaluating public spaces and streetscapes \citep{Mushkani2024AIEDI,Mushkani2026StreetReview}. Reported workflows include iterative refinement of criteria and evaluation protocols.

These cases highlight the need to treat criteria definitions and protocol revisions as versioned artifacts: when criteria change based on participant feedback, the change should be linkable to evidence and to updated tests of change. They also foreground representational questions: group-level differences in preferences are interpretable only when recruitment pathways and role boundaries are documented \citep{Crenshaw1989,Mushkani2026StreetReview}.

\subsection{Case 4: WeDesign+}

WeDesign+ is a participatory text-to-image platform for urban design consultations in which participants co-create images via rapid, collaborative prompting and iterative edits, supported by workshops and expert interviews \citep{Mushkani2025WeDesign}.

The case surfaces a traceability challenge specific to co-creation: prompt changes and negotiation rationales occur in real time, and without structured recording the rationale behind prompt edits and later tool changes can be lost. It also motivates governance and compensation primitives: participants request capability changes (e.g., safety restrictions, multilingual support) and contribute reusable prompt templates and evaluation checks whose value may persist across versions, yet recognition is typically one-off \citep{Mushkani2025WeDesign,GraySuri2019GhostWork}.

\section{Reference schema excerpt}
\label{app:schema}

\begin{table}[t]
\caption{Participation Ledger schema overview (simplified). Fields are illustrative; the excerpt below provides a minimal reference representation.}
\label{tab:schema}
\begin{tabularx}{\linewidth}{@{}lX X@{}}
\toprule
Class & Purpose & Selected required fields \\
\midrule
\texttt{Contribution} & Records participation artifacts such as prompts, labels, rationales, and incident reports. & Stable identifier; contributor role; consent scope; compensation terms; artifact reference (hash or URI); intended use; privacy and retention tags. \\
\texttt{Change} & Records a modification to an AI system artifact (data, prompt library, model adapter, guardrail, deployment config). & Change identifier; references to modified artifacts; \texttt{influencedBy} references; linked tests of change; integrity hash chaining. \\
\texttt{Test} & Records a replayable test of change with input specification and evaluation procedure. & Test identifier; input specification; expected behavior; measurement protocol; provenance links to contributions; evaluation requirements. \\
\texttt{EvaluationRun} & Records execution of tests on a specific artifact version. & Run identifier; evaluated artifact version; test set; results and decision; evaluator role and terms; timestamp; integrity fields. \\
\texttt{Voucher} & Records a governance constraint or authorization on a capability. & Voucher identifier; steward authorization; target capability; conditions and scope; linked evidence; status lifecycle. \\
\texttt{Credit} & Records credit events tied to tests and changes. & Credit identifier; beneficiary (pseudonymous or steward); triggering event; amount/unit; audit link to evaluation run or incident. \\
\bottomrule
\end{tabularx}
\end{table}

This appendix provides a minimal schema excerpt and a minimal example entry, intended to be compilation-safe and illustrative. The schema is presented in prose and an example serialization.

A minimal ledger entry includes an identifier, an entry type, creation timestamp, actor role metadata, consent block, compensation block, a contribution or change summary, linkage fields, and an integrity block containing a hash and optional previous hash pointer and signature. In practice, implementations should adopt JSON-LD contexts aligned with W3C PROV semantics \citep{W3C_PROV_2013,JSONLD_2020} and validate entries with JSON Schema.

\begin{lstlisting}
{
  "id": "pl:contrib:wedesign:prompt:001",
  "type": "Contribution",
  "createdAt": "2025-05-10T14:30:00Z",
  "actor": { "role": "resident", "pseudonym": "P12" },
  "consent": { "status": "granted", "scope": "research+design", "retention": "3y" },
  "compensation": { "model": "honorarium", "amount": 50, "currency": "CAD" },
  "contribution": {
    "kind": "prompt",
    "summary": "Accessible waterfront park with ramps and diverse seating.",
    "artifactRef": "artifact:prompt:sha256:..."
  },
  "links": {
    "influences": ["pl:test:accessibility:001"],
    "evidence": ["pl:evidence:workshoplog:001"]
  },
  "integrity": { "hash": "sha256:..." }
}
\end{lstlisting}

\section{Procurement clause template}
\label{app:procurement}

This appendix provides a template clause intended for adaptation. It is not legal advice.

\begin{lstlisting}
Participation Ledger Requirement (Template).
The Supplier shall provide a Participation Ledger export for each system release and
material update. The export shall (a) include machine-readable participation evidence
fields for contributions used in training, evaluation, and safety controls; (b) provide
traceability links from contributions to changes and to versioned tests of change;
(c) include replayable participatory evaluation tests and evaluation run records for
the delivered release; and (d) document any active Capability Vouchers and their
enforcement status. Failure to deliver an export conforming to the agreed schema
constitutes a material non-conformance.
\end{lstlisting}

\section{Stakeholder interview guide and reporting rubric}
\label{app:interview_guide}

This appendix provides the semi-structured interview guide and reporting rubric referenced in Section~\ref{sec:stakeholder_validation}. It is written to support replication in public-sector and community-governed contexts.

\begin{figure*}[t]
  \centering
  \includegraphics[width=\textwidth]{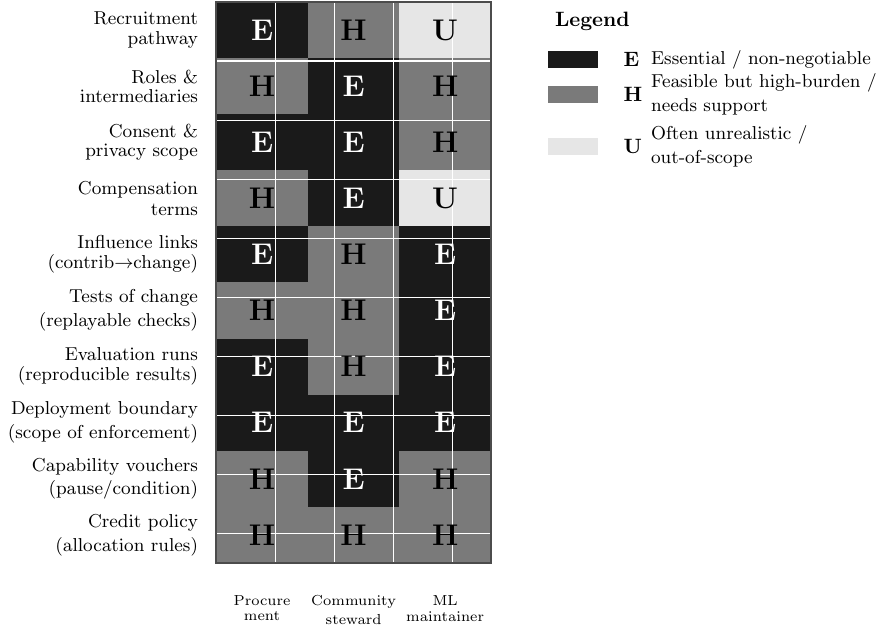}
  \caption{Stakeholder feasibility matrix for coding evidence fields and governance mechanisms in the Participation Ledger workflow. Cells indicate whether each element is treated as Essential (E), feasible but high-burden (H), or often unrealistic/out-of-scope (U) within the stakeholder’s enforcement boundary.}
  \Description{A 10-by-3 matrix with rows listing evidence fields and ledger mechanisms (recruitment, roles and intermediaries, consent and privacy scope, compensation terms, influence links, tests of change, evaluation runs, deployment boundary, capability vouchers, credit policy) and columns listing stakeholder roles (procurement, community steward, ML maintainer). Each cell contains E, H, or U with a grayscale fill, and a legend explains the codes.}
  \label{fig:stakeholder_matrix}
\end{figure*}

\subsection{Sampling frame}
We target 8--15 participants total, spanning three roles:
\begin{itemize}
  \item \textbf{Procurement / contract management}: public-sector staff who write or manage procurement contracts and acceptance criteria for AI systems.
  \item \textbf{Community stewards}: individuals or organizations that recruit participants, manage consent and compensation, and represent community interests in ongoing governance.
  \item \textbf{ML maintainers}: engineers and evaluation and release managers responsible for model updates, evaluation pipelines, and incident response.
\end{itemize}

\subsection{Core workflow walk-through prompts}
We structure interviews around a concrete trigger (e.g., ``a new model version is delivered''; ``a workshop produces prompts/labels''; ``an incident report is filed''). Example prompts:
\begin{itemize}
  \item \textbf{Boundary and levers:} What decisions can you make directly, and what decisions require escalation? What levers (contractual, technical, governance) let you stop, delay, or condition a deployment?
  \item \textbf{Evidence feasibility:} For each evidence field (recruitment, roles, consent scope, compensation, influence links, tests/evaluation runs), what is already captured in your tools? What would be new work, and who would do it?
  \item \textbf{Incentive breaks:} Where would logging and enforcement fail in practice? (e.g., misaligned incentives, missing authority, vendor lock-in, privacy constraints)
  \item \textbf{Minimum viable evidence:} If you could require only 3--5 fields for a first deployment, which would they be and why? What would you explicitly defer?
\end{itemize}

\subsection{Role-specific probes}
\paragraph{Procurement / contract management.} What acceptance tests are already used (security, performance, bias)? Where could participatory tests fit? What contract clauses are feasible (audit rights, deliverables, retention limits)? What evidence is legally or operationally risky to require?

\paragraph{Community stewards.} How is consent tracked today (paper forms, digital systems, verbal consent)? What compensation mechanisms are used, and what constraints exist (timing, taxation, eligibility)? What authority do stewards have to pause deployments or require remediation?

\paragraph{ML maintainers.} How are model updates staged (branching, CI/CD, model registry)? Where do evaluation runs live (CI logs, dashboards, experiment trackers)? What evidence can be auto-populated, and what requires human interpretation (rationales, consent scope)?

\subsection{Reporting rubric}
We code each evidence field and each stakeholder group using the three-level rubric used in Figure~\ref{fig:stakeholder_matrix}:
\begin{itemize}
  \item \textbf{Essential (E):} treated as non-negotiable for accountability or safety.
  \item \textbf{High-burden (H):} feasible but requires new tooling, staffing, or governance support.
  \item \textbf{Unrealistic/out-of-scope (U):} not enforceable within the participant's boundary or would create unacceptable risk/burden.
\end{itemize}

In addition to field-level codes, we capture (i) the specific enforcement levers available (contract clause, CI gate, steward approval), (ii) the point in the workflow where the evidence would be created, and (iii) any identified failure modes (gaming, privacy risks, performative compliance).

\end{document}